# Push-Pull Direct Modeling of Solid CAD Models

Qiang Zou, Hsi-Yung Feng*
Department of Mechanical Engineering
The University of British Columbia
Vancouver, BC
Canada V6T 1Z4

**Abstract**

Direct modeling is a very recent CAD modeling paradigm featuring direct, intuitive push-pull interactions with the geometry of the model to much increase model editing flexibility. The major issue for push-pull direct modeling is the possible inconsistency between the altered geometry of the model and its unchanged topology. The challenge of resolving the geometry-topology inconsistency lies in ensuring that the resulting model remains as a valid solid model and that the model shape follows a continuous change pattern. Although push-pull direct modeling has been implemented in several CAD software packages, robustness towards generating valid modeling results and continuous shape changes still remains an open issue. This paper proposes a systematic method to handle the resolution of any possible inconsistent situation. The method formulates the continuous shape change requirement as successive Boolean operations on the model volume, thereby guaranteeing valid solid models and continuous shape changes. Further, this formulation allows an easy implementation of push-pull direct modeling using existing CAD research and engineering results. In order to show the effectiveness of the proposed method, a software prototype is developed and the modeling results are compared with those of five leading CAD software packages.

**Keywords:** Direct modeling; Push-pull; Solid modeling; Validity; Continuity

## 1. Introduction

Direct modeling is a very recent computer-aided design (CAD) modeling paradigm succeeding the history-based CAD modeling paradigm [1]. This modeling technique aims to increase modeling flexibility and facilitate the modification-based design practice that has been increasingly seen in engineering design [2,3]. The models undergoing modification are usually stored in a neutral CAD model scheme such as IGES and STEP [4]. These models are essentially boundary representation (B-rep) solid models that represent shapes using the boundary between solid and non-solid [5]. In order to achieve efficient modification of such models, direct modeling develops modeling operations that allow designers to interact directly and intuitively with the geometry (boundary) of the model.

In general, any operation serving the purpose stated above can be classified as a direct modeling operation. Among those operations, push-pull operations as depicted in Fig. 1 are the primary operations. For this reason, direct modelers are also called push-pull systems or techniques [6,7]. Push-pull direct modeling allows for direct alterations to positions and/or orientations of boundary faces in a solid model. With it, many geometric details can be easily relocated, deleted or added, which is particularly useful when it is combined with the virtual/augmented reality technique [8].

When push-pull direct modeling is used in real-world mechanical design, it is found that the modeling system may fail to update the model. Quite often, these failures do not arise from the typical numerical issues in many CAD robustness problems [9], but result from the intrinsic ambiguity in push-pull direct modeling. A B-rep solid model consists of information on both geometry and topology, which must be consistent with each other to attain a valid model [5]. When a solid model is push-pulled, only part of its geometry is changed due to the moved boundary faces. This, in many cases, breaks the geometry-topology consistency in the pre-edit model. In consequence, an invalid model (e.g., a self-intersecting model) will be generated. To resolve the geometry-topology inconsistency, there often exist many options (refer to Section 3 for a detailed discussion). Some of them will lead to invalid or non-solid models. Among the valid models, there will only be one in line with the designer's intent. As a result, the resolution process is prone to invalid and unpredictable results. The fundamental challenge for realizing push-pull direct modeling thus lies in the robustness of generating valid and predictable results.

Design intent is generally too complicated to infer exactly by a computer [10]. Fortunately, it has been shown in [11,12] that much, if not all, of the unpredictable modeling behavior in CAD can be avoided if the model shape follows a continuous change pattern. The problem of generating valid and predictable modeling results is thus narrowed as generating valid models with continuous shape changes. This formulation is believed to serve as the first step towards fully predictable push-pull direct modeling systems. Rather than adopting a heuristic approach, which can provide acceptable results in many but not all of the potential cases, this work develops a systematic method for push-pull direct modeling to attain guaranteed model validity and continuous shape change.

* Corresponding author. Tel.: +1-604-822-1366; fax: +1-604-822-2403. E-mail: feng@mech.ubc.ca

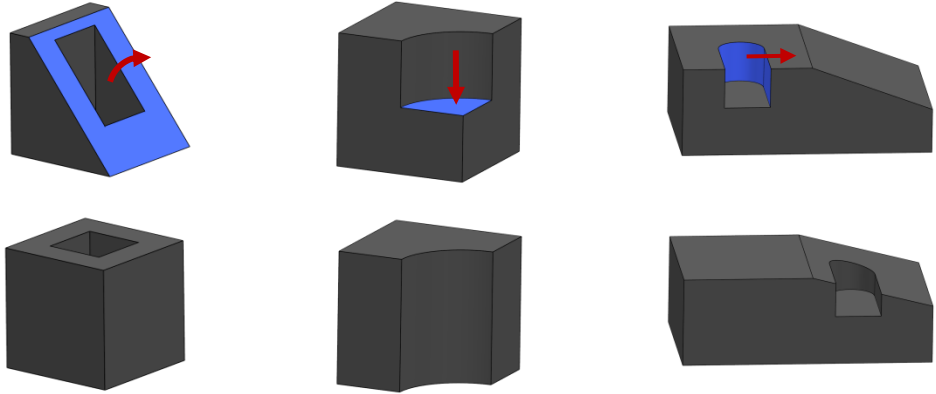

Figure 1   Push-pull direct modeling examples.

## 2. Related work

Push-pull direct modeling has a predecessor from the 1970s and it was then called tweaking [13,14]. It did not allow violation to the pre-edit topology; therefore, there was no robustness issue. Modern push-pull direct modeling discards this restriction in order to achieve much improved modeling flexibility. However, the underlying robustness issue has rarely been noted and studied by the academic community (partially because direct modeling is relatively new) and only partially solved by industry. A few CAD software packages are being marketed with push-pull direct modeling. Without access to how they implement the push-pull direct modeling, an in-depth review and evaluation cannot be performed. Nevertheless, it appears that heuristic approaches have been employed due to the varying modeling results observed: some results are good but others are not.

In the literature, Rossignac [15] briefly discussed the conversion of translational push-pull operations to face extrusion operations in the assessment of the feature modeling technique. However, no systematic methodology was explained. Woo and Lee [16], Kim and Mun [17] and Fu et al. [18] proposed to interpret push-pulling a face as push-pulling the volumetric feature containing the face. This line of methods clearly ensures model validity but, they are used to change the volume of a feature rather than individual faces. Hence, their application in model modification is restricted. That is, the work of [16,17] allows for translations and/or rotations of the volumetric feature as a whole but, editing the feature itself is not available. The work of [18] further allows the designer to parametrically edit the volumetric feature but, making topological edits to the volumetric feature is not available. Lipp et al. [19] developed a bunch of heuristics to handle the geometry-topology inconsistency resulting from push-pulling polygonal mesh models. However, the heuristics were developed specifically for models composed only of planar and non-holed boundary faces, which merely make up a small portion of solid models.

To date, the documented research study on the robustness of push-pull direct modeling is insufficient. The modeling software vendors seem to have good grasp on how to implement the push-pull direct modeling using their own proprietary methods. Unfortunately, their methods are not complete. A robust method for implementing push-pull direct modeling is to be presented in this paper. It begins in the next section with a detailed analysis on push-pull direct modeling, explaining the key issues that have to be resolved. Section 4 outlines the proposed continuity-based method with the details given in Section 5. Performance comparison of the proposed method with several leading CAD software packages is found in Section 6, followed by conclusions in Section 7.

## 3. Issues in push-pull direct modeling

A solid is defined as a subset of $R^3$ that is bounded, closed, regular and semi-analytic, often referred to as an r-set [5]. This definition encompasses the common attributes of most mechanical parts [20]; however, it allows non-manifold models. For this reason, a solid is preferred to be an r-set with a 2-manifold boundary [21]. A solid model is a computational or digital representation of a solid. The B-rep scheme represents a solid as a collection of boundary faces [5]. Data stored in a B-rep model are usually classified into geometry and topology. The geometry refers to the numerical data, including surfaces, curves and points. The topology refers to the symbolic data, i.e., connections between the geometric elements. The geometry and topology are interdependent because the topology defines how a surface in the geometry is trimmed to a boundary face [22].

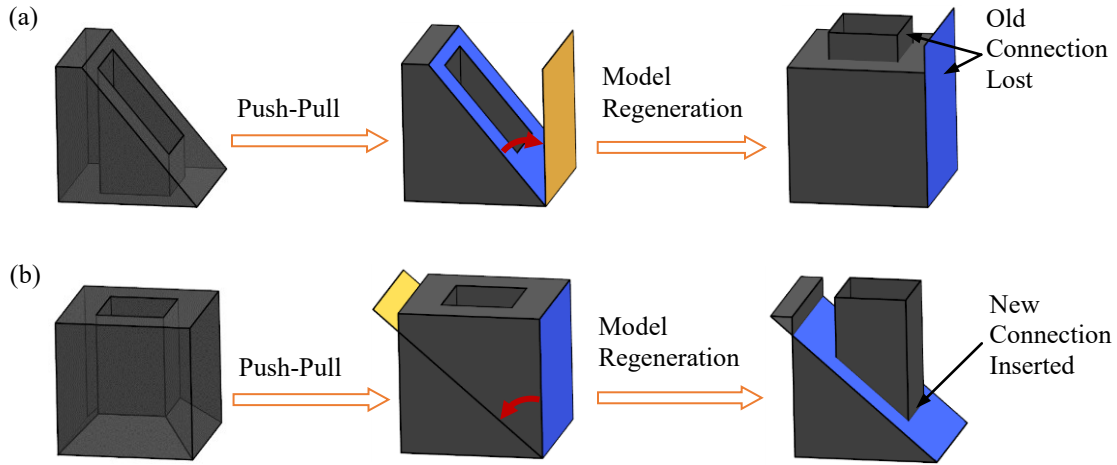

Figure 2   Examples of model regeneration failure types.

While any solid can be represented by the B-rep scheme, a B-rep model does not always give a solid model. A B-rep model is a solid model if the following conditions are satisfied [5,21,23,24]:

1. A face is a bounded, regular and semi-analytic subset of an at least piecewise analytic surface in $R^3$ and the face interior must be connected;
2. Each edge is shared by exactly two faces;
3. Faces around each vertex form a closed fan; and
4. Faces may intersect only at common edges or vertices.

Essentially, Condition 1 requires faces to be well-bounded, meaning each face boundary is not open or intersected [25]. Conditions 2 and 3 ensure manifoldness along each edge and around each vertex. Condition 4 ensures manifoldness of the face interior.

### 3.1. Boundary model regeneration

When push-pull direct modeling is applied to a B-rep solid model, only the target locations and orientations of the push-pulled faces are specified. The associated changes made to the boundaries of the push-pulled faces as well as those of the other affected faces are not quite known. Hence, every push-pull move needs to be followed by a boundary regeneration process for the push-pulled solid model. This regeneration is readily performed through intersecting the post-edit surfaces with respect to the pre-edit topology.

Model regeneration may fail to output a valid solid model due to the inconsistency between the altered model geometry and pre-edit model topology. The geometry-topology inconsistency results from either losing an old connection or inserting a new connection, as shown in Fig. 2. Losing an old connection means that there is no solution to the intersection of the involved surfaces or curves. These faces are then open (ill-bounded), violating the solid model validity Condition 1. Inserting a new connection means that there exist extra intersections within the face, violating the validity Condition 4 and again, leading to an ill-bounded face. A special inconsistency case also exists. For a push-pulled face with a vertex incident to four or more faces, it is possible that faces are well-bounded but non-manifoldness occurs on edges or vertices, violating the validity Conditions 2 and 3. However, such a design is not common in mechanical parts due to the much increased manufacturing cost. This work thus does not deal with such cases.

After model regeneration, if there is no ill-bounded face, a valid model is output and nothing further needs to be done. If not, the involved geometry-topology inconsistency needs to be resolved. This is the core task to implement push-pull direct modeling but current CAD software packages are unable to handle it satisfactorily. For the simple cases in Fig. 2, five leading modeling software packages failed for the case of Fig. 2a and four of them failed for the case of Fig. 2b.

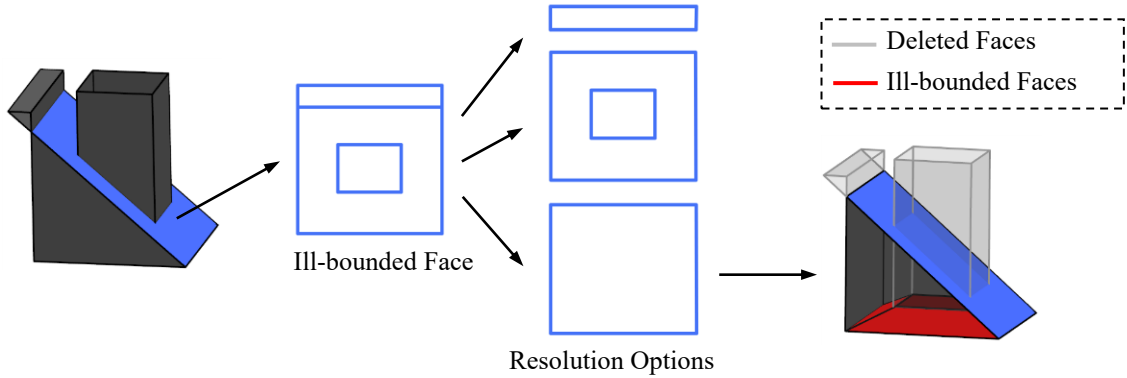

Figure 3   An incorrectly selected resolution option leading to an invalid model.

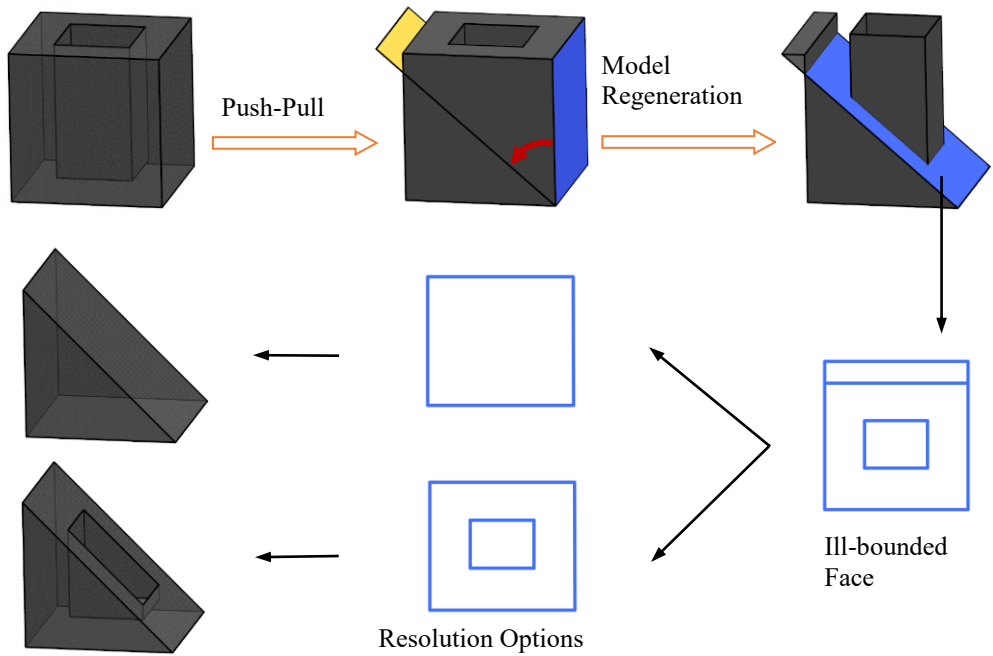

Figure 4   Different applicable resolution options leading to varied model shapes.

### 3.2. Robust resolution issues

In resolving the geometry-topology inconsistency stated above, the applied resolution method should always output valid models and the models should follow a continuous change pattern. However, this is not straightforward due to the intrinsic ambiguity in the inconsistency resolution.

It is challenging to have a validity-guaranteed resolution result because there exist many possible resolution options. For example, Fig. 3 shows three sample options for resolving an ill-bounded face for the case of Fig. 2b. These options together with those due to other ill-bounded faces correspond to multiple outcomes for the whole model. Further, to ensure that any edge on a resolved face is shared with a neighboring face to satisfy the manifoldness condition, the neighboring faces has to be well-bounded as well. Thus, decision-making in the inconsistency resolution is error-prone. For instance, if the bottom option for the case of Fig. 3 is chosen, all the faces depicted as gray should not exist. These faces will thus be deleted. This in turn makes the previously well-bounded faces that are depicted as red ill-bounded, indicating a failed resolution.

It is equally challenging to attain continuous model shape changes even the resolution process successfully avoids invalid results. To maintain a manifold solid model, a choice made locally to a face affects its neighboring faces, and so forth. A local choice thus has a global effect on the whole model. As a result, decision-making in the inconsistency resolution is very subtle. For example, Fig. 4 shows that two different local choices can lead to two completely different outcomes. Therefore, it is hard to make appropriate local choices so that the results follow a continuous change pattern. To date, no systematic methods for resolving this issue have been reported; commercial CAD software packages are not able to handle this issue robustly. Only three out of the five tested packages successfully gave the bottom model shape in Fig. 4. The other two packages generated the model shape with the blind hole disappeared. The three modeling packages that succeeded in this case failed to generate desirable results for the cases in Section 6.

## 4. The proposed methodology

In order to address the robust resolution issues of push-pull direct modeling, a systematic geometry-topology inconsistency resolution method is needed. To achieve this goal, this work first formalizes the continuity requirement and then, based on the results, formulates the decision-making in the inconsistency resolution as successive Boolean operations on the model volume. This Boolean operation based formulation clearly guarantees valid results. It also yields a continuous shape change since the formulation is rooted in the formalization of continuity.

### 4.1. Principle of continuity

Conceptually, the principle of continuity states that an infinitesimal change in a modeling operation will only yield an infinitesimal change in the resulting model [11,12]. An infinitesimal change in push-pull direct modeling means an infinitesimal change in the push-pull parameters such as the translated distance and/or rotated angle. An infinitesimal change in the resulting model refers to an infinitesimal change to some attribute of the model. More precisely, let the push-pull direct modeling be represented by the rigid transformation matrix applied by the user, denoted as $T(t), t \in [0,1]$, and the model variation over the push-pull process be represented as $M(t), t \in [0,1]$. The principle of continuity can then be formulated as:

$$f\big(M(t+\varepsilon), M(t)\big) \to 0 \quad \text{when} \quad \varepsilon \to 0 \tag{1}$$

where $f(M_1, M_2)$ is a function quantifying the difference between models $M_1$ and $M_2$ with respect to a certain model attribute. To determine an applicable model attribute to be used in the function $f$, model face area and model volume that are directly related to visual perception are potential candidates. However, model faces may cancel each other out during a push-pull move, resulting in a sudden and discontinuous change for the model face area. Fig. 5 depicts a common example where push-pulling the blue face immediately after the orange location will cause face F1 to be deleted suddenly, leading to a discontinuous change in the model face area. As a result, the model face area is not an applicable model attribute to be used in $f$ for push-pull direct modeling. Rather, the function $f$ is to be formulated in this work using the model attribute of volume to maintain continuous changes.

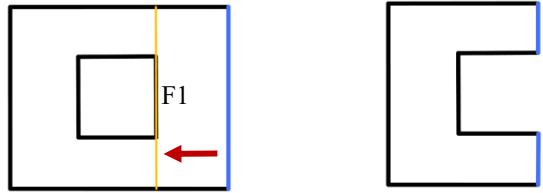

Figure 5   A common push-pull operation leading to a sudden deletion of face F1.

### 4.2. Topology change point based decomposition

Despite a push-pulled face is being moved continuously, the involved inconsistency resolution is in fact to be handled in a discrete manner. The push-pull process can be classified into different phases and it is transitioning from one phase to the next by going through inconsistency resolution. Consider, for example, the push-pull direct modeling depicted in Fig. 6a. Let $t$ denote an intermediate value within the push-pull domain [0,1]. If $t$ is relatively small, it is found that inconsistency resolution is not needed since the regenerated model remains valid and consistent with the principle of continuity (Fig. 6b). Increasing $t$ to a certain value would necessitate inconsistency resolution. Specifically, as depicted in Fig. 6c, inconsistency resolution at $\tau - \varepsilon$ is still not needed but it is needed at $\tau + \varepsilon$ to resolve the invalid model shown in Fig. 6d. As can be seen, it is the crossing of the point $\tau$ that the need of inconsistency resolution emerges and that changes to the model topology are needed. The more of such points a push-pull move needs to cross, the more inconsistency resolution is needed. Due to the complication associated with crossing each such point, it is less likely that a valid result with a continuous shape change can be attained when many of such points are to be crossed. These points are referred to in this work as topology change points (TCPs). More formally, for an infinitesimal perturbation $\varepsilon$ and the models at $\tau - \varepsilon$ and $\tau + \varepsilon$ have different topologies, the point $\tau$ is deemed a TCP.

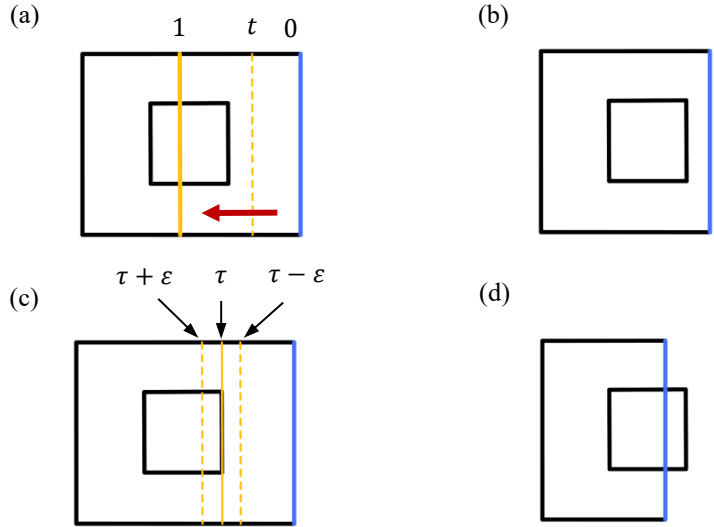

Figure 6    Topology change point: (a) typical push-pull operation; (b) regenerated model at $t$; (c) crossing a TCP at $\tau$; and (d) regenerated invalid model at $\tau + \varepsilon$.

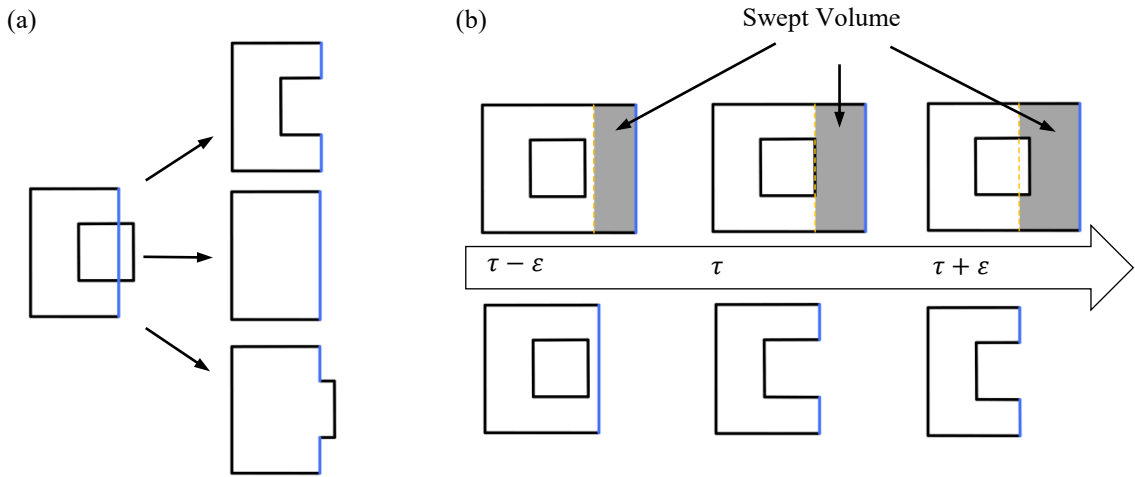

Figure 7    Crossing a TCP: (a) resolution options; and (b) continuous $M(t)$ attained from a swept volume based $\Delta M(t)$.

With the TCPs in place, push-pull direct modeling can be applied progressively. Consider a push-pull move crossing only one TCP, like the case of Fig. 6a. Instead of resolving the geometry-topology inconsistency at the end of the push-pull move, the inconsistency formed immediately after the TCP is resolved first, resulting in an intermediate model. The rest of the push-pull direct modeling is then applied to the intermediate model. In other words, the original push-pull move is decomposed into two sequential, smaller push-pull moves. The first applies the transformation $T(\tau + \varepsilon)$ to the original model $M(0)$ where $\tau$ denotes the TCP. The second applies the remaining transformation $T(1)T(\tau + \varepsilon)^{-1}$ to the intermediate model $M(\tau + \varepsilon)$. This division is beneficial because inconsistency resolution at $\tau + \varepsilon$ is simple since a valid reference model $M(\tau - \varepsilon)$ is close by and the principle of continuity can thus be exploited while inconsistency resolution at the end of the push-pull move is not under such an advantageous situation.

For general push-pull direct modeling with multiple TCPs, a sequence of intermediate models right after each TCP is to be generated. With this sequential approach, the original one-time decision-making for inconsistency resolution is converted into a sequence of smaller and simpler decision-makings. Since each smaller decision-making maintains a continuous change to the model, the principle of continuity is satisfied for model $M(t)$ throughout the push-pull process.

### 4.3. Boolean operation based decision-making

Resolution options emerge whenever a TCP is crossed. For example, Fig. 7a shows three sample options for the case depicted in Fig. 6. The conceptually simplest way to satisfy the principle of continuity is to set a small $\varepsilon$ and choose the resolution option that gives the smallest volumetric difference from the model $M(\tau - \varepsilon)$ and the top option in Fig. 7a will be chosen. However, this requires an exhaustive list of resolution options to be available for comparison. A more effective way is to use Boolean operations to get to the intended resolution option directly.

It should be pointed out that as the model attribute of volume has been chosen to formulate model difference, a continuous $M(t)$ means that changes made to the model volume $\Delta M(t)$ is also continuous. Consider, for example, the case depicted in Fig. 6a. The model volume change at $t$ is given by:

$$\Delta M(t) = M(0) - M(t) \tag{2}$$

where "−" denotes the regularized Boolean difference operator. At $\tau - \varepsilon$, the model volume change is simply the volume swept by the push-pulled face from 0 to $\tau - \varepsilon$ as shown in Fig. 7b. By continuing this sweeping across the TCP, a continuous model volume change is achieved over the domain $[\tau - \varepsilon, \tau + \varepsilon]$. Refer to Fig. 7b for an overall illustration of this process. With the model volume change $\Delta M(\tau + \varepsilon)$ known, the model $M(\tau + \varepsilon)$ can then be obtained by a simple Boolean difference operation:

$$M(\tau + \varepsilon) = M(0) - \Delta M(\tau + \varepsilon) \tag{3}$$

In the above equation, it is evident that if $\Delta M(t)$ is negative, material/volume will be added to the original model $M(0)$.

In the discussion above, a small perturbed value $\varepsilon$ is utilized. In fact, the swept volume based $\Delta M(\tau + \varepsilon)$ can be divided into two parts: one from 0 to $\tau$ and the other from $\tau$ to $\tau + \varepsilon$. The latter part is then combined with the swept volume based $\Delta M(t)$ from $\tau + \varepsilon$ to 1. This divide-combine procedure should also be applied to a push-pull move with multiple TCPs. Hereafter, the model volume change $\Delta M(t)$ from a TCP to the next TCP is referred to in this work as an auxiliary model. Combining the TCP based decomposition procedure and the Boolean operation based decision-making yields the proposed continuity-based method, which consists of the following three basic steps:

| **Algorithm 1**: Continuity-Based Method |
| --- |
| **Repeat** (initially, $i = 0$ and $t_0 = 0$) |
| I.    Detect next TCP, represented as $t_{i+1}$ |
| II.   Construct the auxiliary model $\Delta M_i^{i+1}$ for this iteration |
| III.   Update the model $M(t_{i+1}) \leftarrow M(t_i) - \Delta M_i^{i+1}$ |
| **Until** $t_{i+1} = 1$ |

## 5. Method details

The previous section outlines the proposed continuity-based method. This section gives details about Step I (TCP detection) and Step II (auxiliary model construction) of Algorithm 1 in push-pulling a single face (Subsections 5.1 and 5.2) as well as multiple faces concurrently (Subsection 5.3).

### 5.1. Topology change point detection

TCPs correspond to critical points in a push-pull move at which ill-bounded faces are formed. Getting the next TCP is effectively to predict all TCPs and then choose the closest one. The key to this process is to have an exhaustive list of predicted TCPs. To achieve this, all model faces are divided into two groups: the inner group consisting of the push-pulled face and its neighboring faces, and the outer group consisting of the remaining faces. Then, we can list all the possible ways the inconsistency could occur, as shown in the left column of Table 1. Each of the possible ways further has two inconsistency types according to the sources of the inconsistency. Three out of the six situations, as indicated by cross marks in Table 1, are impossible because all faces except for the push-pulled face are set to be stationary and stationary faces cannot form ill-bounded faces by themselves. Fig. 8 shows the examples of the other three possible situations as indicated by check marks in Table 1.

To detect the next TCP, we basically check the three possible situations listed in Table 1 in a sequential manner and then select the one with the smallest push-pull parameter, as shown in Algorithm 2. The three situations obviously have different natures and thus, their checking methods are made different as follows. For the first situation of inserting new connections on inner-group faces only, the checking is performed through detecting surface-surface intersections (Lines 2-9). More specifically, a push-pull move will shrink or expand inner-group faces on their respective underlying surfaces and hence, the potential next TCP corresponds to a certain intersection between the underlying surfaces of the inner-group faces, as shown in Fig. 9a.

Table 1   Geometry-topology inconsistency types.

| Faces involved in inconsistency | Inconsistency types | |
|---|---|---|
| | Inserting new connections | Losing old connections |
| Within inner group | ✓ (Fig. 8a) | ✓ (Fig. 8c) |
| Within outer group | ✗ | ✗ |
| Between inner and outer groups | ✓ (Fig. 8b) | ✗ |

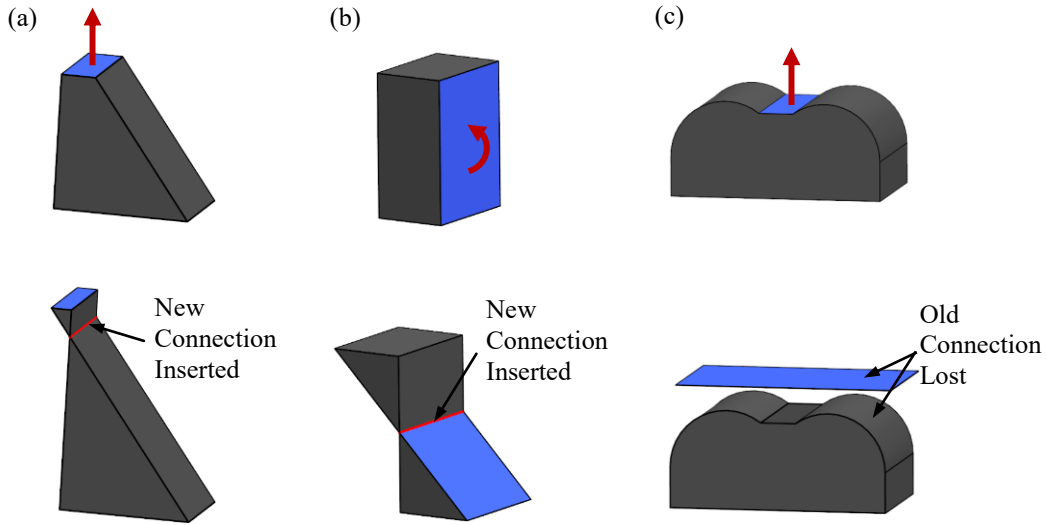

Figure 8   Illustrations of geometry-topology inconsistency types.

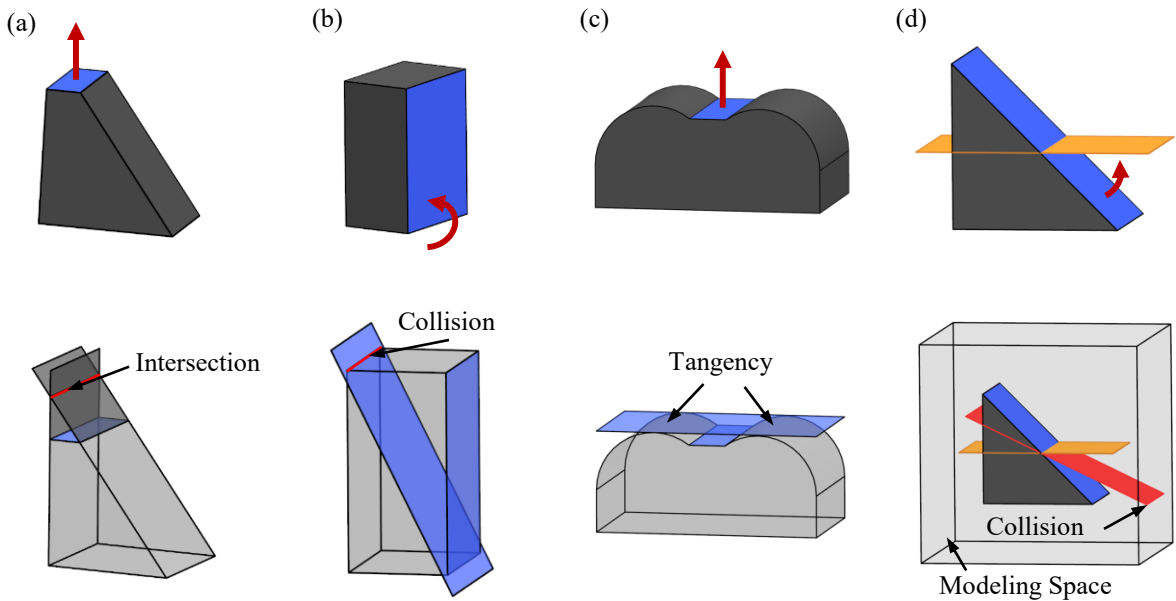

Figure 9   Candidates for the next TCP.

For the second situation of inserting new connections between inner and outer group faces, the checking is done by detecting surface-face collisions (Lines 10-14).  That is, the push-pulled face will be the first face in the inner group to meet the outer-group faces since other faces are stationary. Thus, the corresponding checking is to determine the first contact or collision between the push-pulled surface and outer-group faces, as shown in Fig. 9b. It should be noted here that collisions occurring outside the (dynamic) boundary of the push-pulled face should not be considered applicable.

For the third situation of losing old connections on inner-group faces only, the checking is made by surface-surface tangencies or face-workspace collisions (Lines 15-21). To be precise, this situation indicates that the push-pulled surface no longer intersects its neighboring surfaces. Surface-surface intersection is lost either due to one surface moves beyond the size limit of the other surface, or to two surfaces become parallel and intersect at infinity. For the first reason, the push-pulled surface become tangent to a neighboring surface signals that the related intersection is about to be lost (Fig. 9c). For the second reason, the model becomes infinitely large with the push-pull move. However, a design model undergoing modification cannot be made larger than the modeling space. Hence, the geometric event signaling this situation is the collision between an inner-group face and a pre-set box representing the modeling space (Fig. 9d).

It is obviously that Algorithm 2 relies on the algorithms of detecting intersections, collisions and tangencies between surfaces or faces; fortunately, these algorithms are readily available and well established in computational geometry [26,27].

---

**Algorithm 2**: TCP Detection

**Input:** $M, F_{pp}$ −The model and the push-pulled face

**Output:** $t$ − Next TCP

1.   $t \leftarrow 1$ // push-pull parameter, 1 means the end of the push-pull
2.   $F_{nei} \leftarrow$ GetNeighboringFaces($F_{pp}$)
3.   **for** each face pair $(f_1, f_2)$, $f_1, f_2 \in F_{nei}$ **do**
4.       $s_1, s_2 \leftarrow$ GetUnderlyingSurface($f_1, f_2$)
5.       **if** isIntersect($s_1, s_2$) = $TRUE$ **then**
6.          $t_p \leftarrow$ GetIntersectionPoint($s_1, s_2$)
7.          **if** $0 < t_p < 1$ **then** $t \leftarrow$ Min($t, t_p$)
8.       **end if**
9.   **end for**
10.  $S_{pp} \leftarrow$ GetUnderlyingSurface($F_{pp}$)
11.  **for** each outer-group face $f \in M$ **do**
12.     $t_p \leftarrow$ GetCollisionPoint($f, S_{pp}$)
13.     **if** $0 < t_p < 1$ and in $F_{pp}(t_p)$ **then** $t \leftarrow$ Min($t, t_p$)
14.  **end for**
15.  **for** each face $f \in F_{nei}$ **do**
16.     $s \leftarrow$ GetUnderlyingSurface($f$)
17.     $t_p \leftarrow$ GetTangencyPoint($s, S_{pp}$)
18.     **if** $0 < t_p < 1$ **then** $t \leftarrow$ Min($t, t_p$)
19.     $t_p' \leftarrow$ GetCollision($f, ModelingSpace$)
20.     **if** $0 < t_p' < 1$ **then** $t \leftarrow$ Min($t, t_p'$)
21.  **end for**
22.  **Return** $t$

---

### 5.2. Auxiliary model construction

Auxiliary model construction is to get the volume between a TCP and the next one. According to the TCP definition, no outer-group faces will be involved in the push-pull move before the next TCP. This means that faces involved in the auxiliary model construction are restricted to inner-group faces. The volume swept by the push-pulled surface and bounded by the neighboring surfaces thus gives the auxiliary model. The construction of the B-rep model for this volume effectively follows this sweeping process, as detailed in Algorithm 3 and illustrated in Fig. 10a. First, a tube-like shell out of the neighboring faces is constructed by extending the neighboring faces on their underlying surfaces towards the push-pulled face (Line 2). Each face in this shell is then trimmed at the start and target location and orientation of the push-pulled surfaces (Lines 5-9). The trimmed shell along with the start and target faces form the desired B-rep model (Lines 8, 10 and 11).

It should be noted that when the start and target surfaces intersect within the tube-like shell, a self-intersecting auxiliary model would be generated (Fig. 10b). For such a situation, the resulting volume should be decomposed along the intersection into distinct parts as illustrated in Fig. 10b. The additive part is then added to the original model and the subtractive part subtracted from the original model.

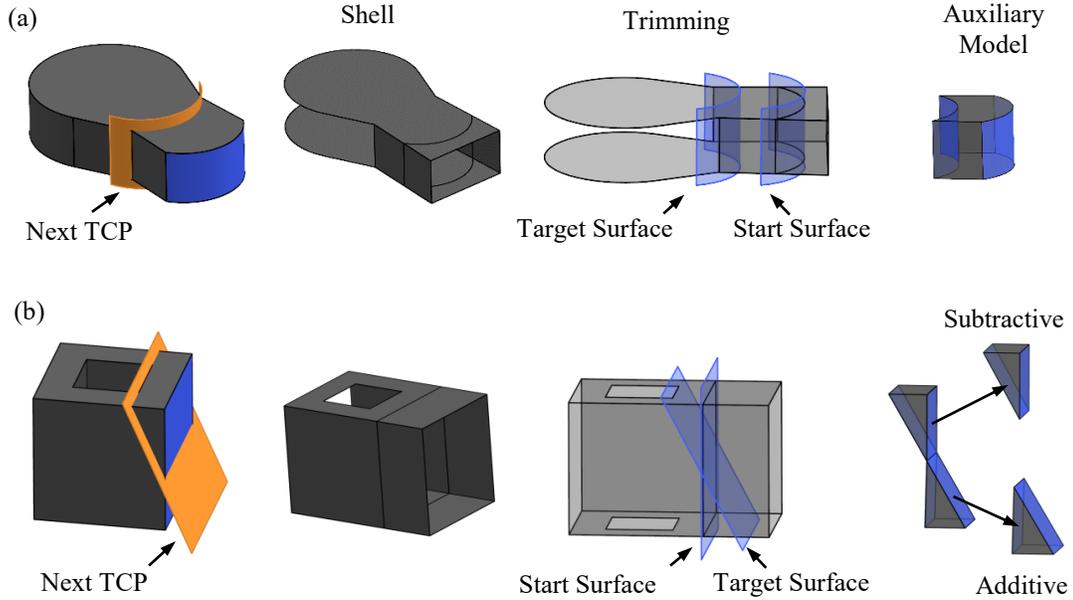

Figure 10   Illustration of auxiliary model construction.

| **Algorithm 3**: Auxiliary Model Construction |
|---|
| **Input:** $F_{nei}, F_{pp}, T$ − Neighboring faces, push-pulled face and the rigid transformation from a TCP to the next one |
| **Output:** $A$ − Auxiliary model |
| 1.    $A \leftarrow \emptyset$ // array of boundary faces of auxiliary model |
| 2.    $F_{nei} \leftarrow \text{ExtendFaces}(F_{nei})$ |
| 3.    $S_s \leftarrow \text{GetUnderlyingSurface}(F_{pp})$ // start surface |
| 4.    $S_t \leftarrow \text{TransformSurface}(S_s, T)$ // target surface |
| 5.    **for** each face $f \in F_{nei}$ **do** |
| 6.        $f \leftarrow f − \text{HalfSpaceFrom}(S_s)$ // Trim face with start surface |
| 7.        $f \leftarrow f − \text{HalfSpaceFrom}(S_t)$ // Trim face with target surface |
| 8.        add $f$ to $A$ |
| 9.    **end for** |
| 10.  add $f \leftarrow \text{TrimSurfaceWithNeighborFaces}(S_s, A)$ to $A$ // start face |
| 11.  add $f \leftarrow \text{TrimSurfaceWithNeighborFaces}(S_t, A)$ to $A$ // target face |
| 12.  **Return** A |

*5.3. Push-pulling multiple faces concurrently*

Concurrent push-pulling of multiple faces can be done by applying the method presented previously to each individual push-pulled face. In practical implementation, a direct application may however produce the issues of missed volume or overlapped volume if the push-pulled faces are adjacent. Consider the common V-slot mechanical part in Fig. 11. A missed volume between the individual auxiliary models is generated by the two adjacent faces. To prevent such a missed volume, the involved auxiliary models are to be modified slightly. As shown in Fig. 12, the missed volume is prevented by preserving the edge-adjacency of the push-pulled faces and merging the edge-adjacent faces. It is simple to see that, if multiple faces incident to a vertex are push-pulled, merging edge-adjacent faces automatically retains the vertex-adjacency of the push-pulled faces.

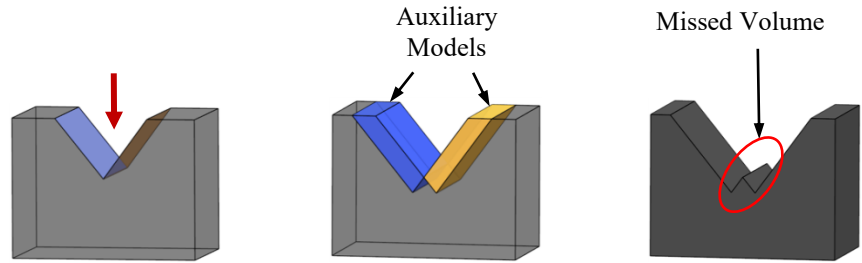

Figure 11   Issue of missed volume produced in concurrent push-pulling of adjacent faces by directly applying the single-face push-pulling method.

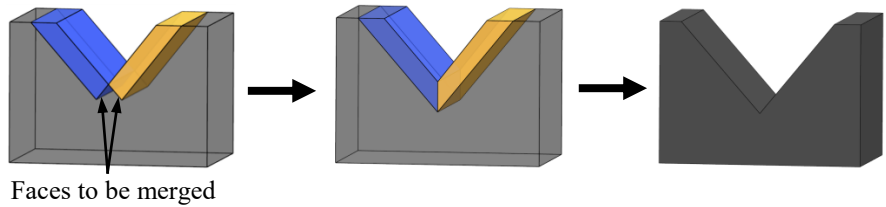

Figure 12   Preventing missed volume between adjacent auxiliary models by merging edge-adjacent faces.

It is known that the result of Boolean operations is order-dependent when the operand models are volumetrically overlapped [5]. Consider again the V-slot model in Fig. 11. If we push-pull toward a horizontal direction, two different modeling results will be generated, as depicted in Fig. 13. To resolve this issue, a subdivision method is to be used. Basically, the overlapped volume is to be isolated and treated separately. Let the two volumetrically overlapped auxiliary models be denoted as $A$ and $B$. They can be subdivided into three parts: the intersection part $A \cap B$ (the overlapped volume) and the remaining parts $A - (A \cap B)$ and $B - (A \cap B)$. The remaining parts $A - (A \cap B)$ and $B - (A \cap B)$ are then added to or subtracted from the original model accordingly. If $A$ and $B$ are both addition (or subtraction), the intersection part $A \cap B$ is then to be added (or subtracted). If $A$ and $B$ are different (one addition and the other subtraction), the intersection part $A \cap B$ is then to be dropped due to the cancellation of the involved addition and subtraction. Fig. 14 shows the application of this subdivision method to the case in Fig. 13.

## 6. Implementation and case studies

### 6.1. Implementation

The push-pull direct modeling method presented previously has been implemented using C++ on top of the geometric modeling kernel Open CASCADE 7.0 [28]. The graphical user interface module (Fig. 15) was developed by using QT 5.7, taking the geometry processing and rendering framework OpenFlipper [29] as a reference source. To push-pull a solid model like the one in Fig. 15, the user presses the left button in the push-pull toolbox to activate the push-pull direct modeling function and then selects the faces of interest. When the selection is done, a graphical push-pull handle (on top of the blue faces in Fig. 15a) pops up which allows the user to move the faces around. More specifically, the user clicks the two rectangular sub-handles labelled as 3 and 4 in Fig. 15a and moves the mouse to translate and/or rotate the push-pull handle (and therefore the selected faces). The rotation axis can be controlled by moving the rectangular sub-handle that is labelled as 5 in Fig. 15a. During the moving of the push-pull handle, the distance of translation and/or angle of rotation will be displayed besides the corresponding sub-handles. When the moving is done, the method presented previously will update the model according to the specified push-pull parameters. For example, Fig. 15b shows the updated model for translating the blue faces downwards by $20mm$.

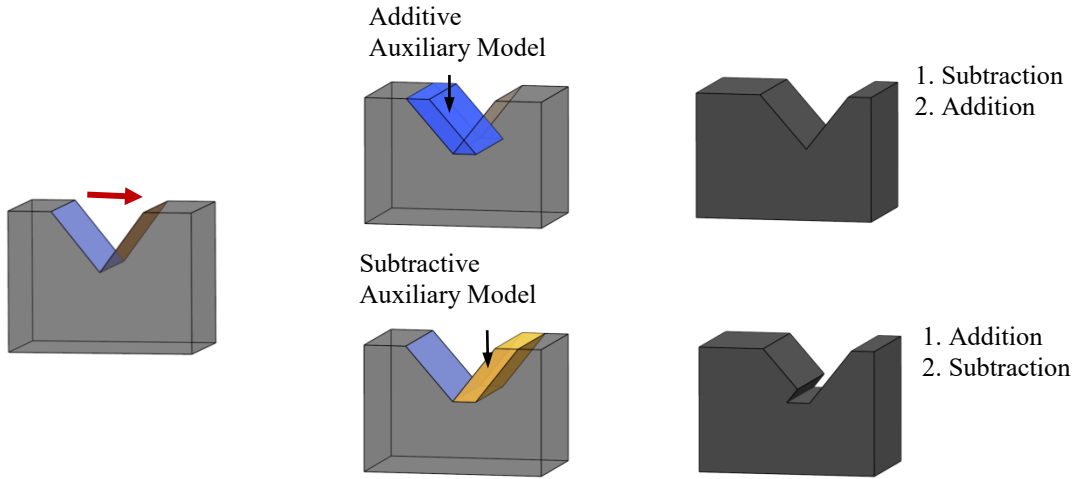

Figure 13   Order-dependency of Boolean operations for overlapped volumes.

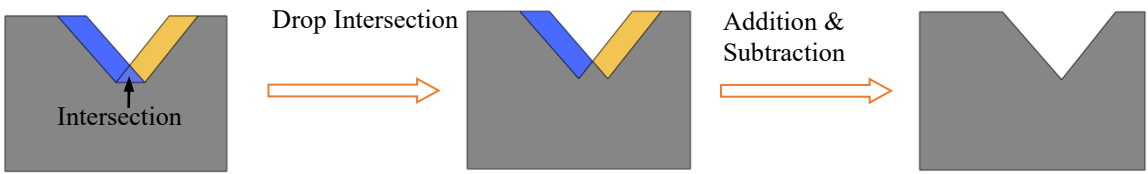

Figure 14   Application of the three-part subdivision method to the case in Fig. 13.

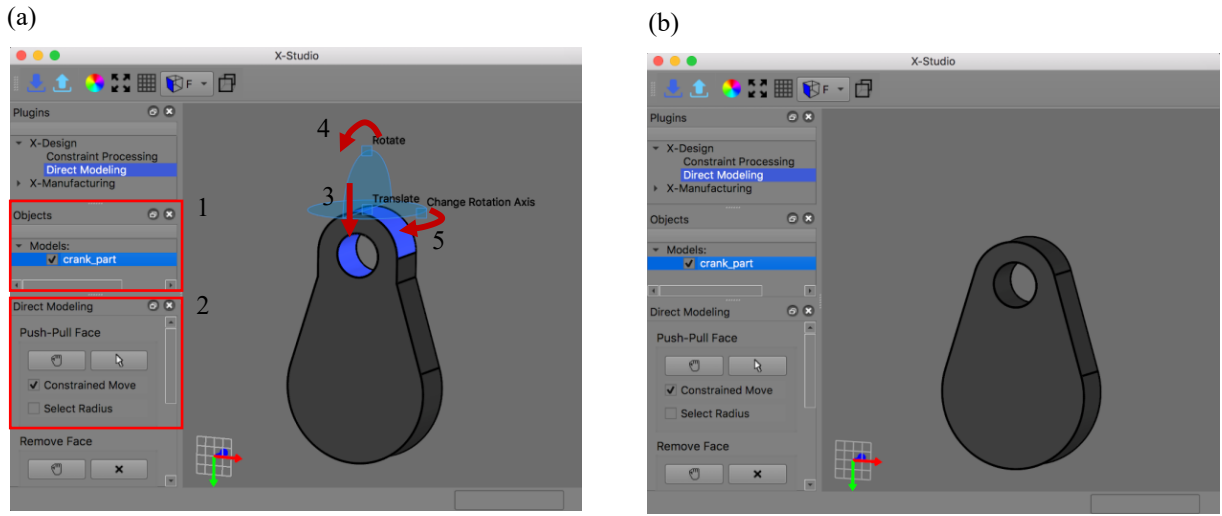

Figure 15   Graphical user interface (a) of the push-pull direct modeling system (labels are: 1. opened model, 2. push-pull toolbox, 3. translate faces, 4. rotate faces, 5. change rotation axis) and (b) modeling result of translating blue faces.

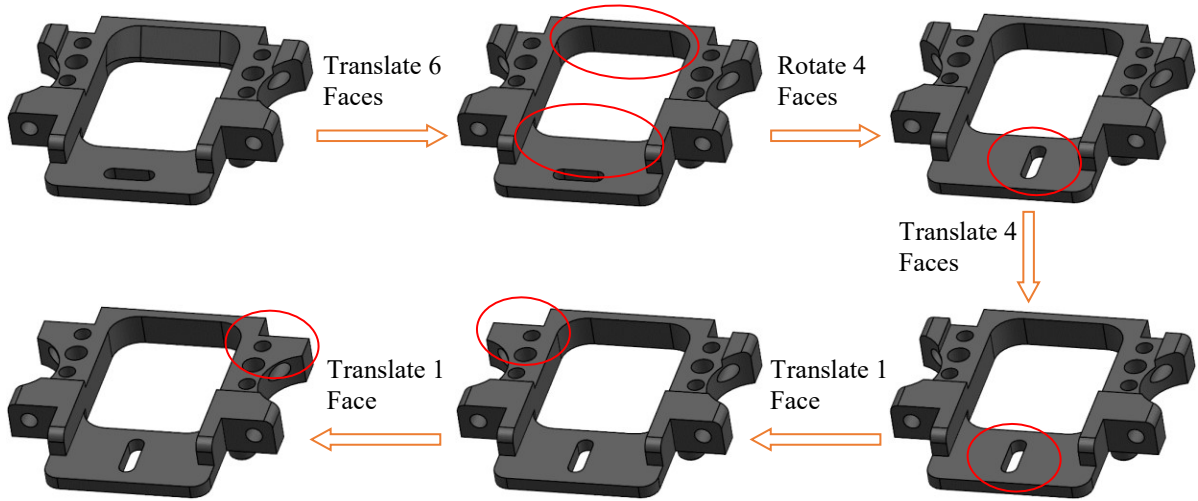

Figure 16   Modeling results of push-pulling the gear box mount model (circles indicate changed parts).

## 6.2. Case Studies

Three case studies have been conducted to show effectiveness of the proposed method. All the part models in these case studies were obtained from the GrabCAD part library (https://grabcad.com/library). The first case study involved a gear box mount part model and demonstrated a comprehensive push-pull direct modeling situation with multiple push-pull edits and both translational and rotational push-pull types (Fig. 16). The purpose of this case is to show that the proposed method as a whole can effectively modify existing models. The second and third case studies (Fig. 17) were intended to show and analyze the respective effectiveness of the two constituent modules of the proposed method: TCP based decomposition and Boolean operation based decision-making. The second case study considered push-pulling a vice base part model and the third case study involved a crank part model. If the model shape changes follow a continuous change pattern, the models depicted in the right column of Fig. 17 are to be expected. The second case study has multiple TCPs (top row of Fig. 18) and therefore, it can be used to illustrate the function of TCP based decomposition. The third case study has one TCP (top row of Fig. 19) and thus, the influence of TCP based decomposition is not significant; Boolean operation based decision-making becomes dominant. This case can thus be used to analyze the function of Boolean operation based decision-making. The modeling results of the proposed method have been compared with those of five leading commercial mechanical CAD software packages: ANSYS SpaceClaim 19, Siemens NX 11, PTC Creo Elements/Direct Modeling 19, SolidWorks 2018, and Autodesk Inventor 2018, as shown in the bottom rows of Fig. 18 and Fig. 19.

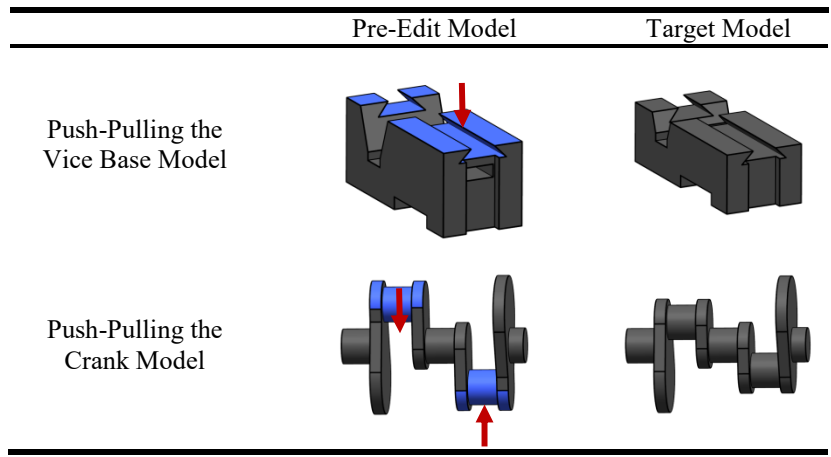

Figure 17   Models chosen for the comparison case studies.

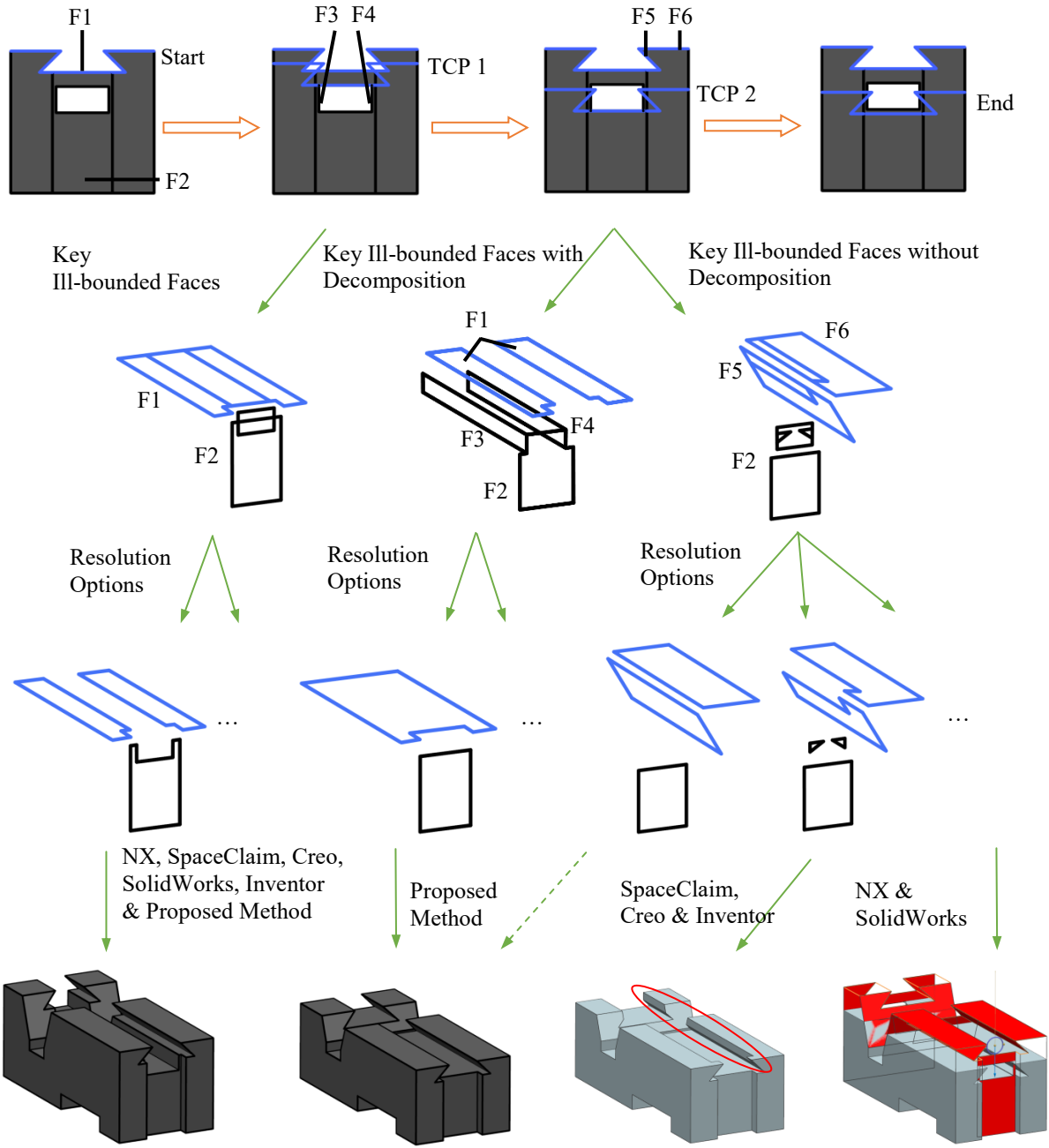

Figure 18   Comparison of modeling results for push-pulling the vice base model.

## 6.3. Discussion

The involved push-pull move in the second case study needs to cross two TCPs.  One is the bottom face of the dovetail (labelled as F1) crossing the top face of the hole below the dovetail.  The other is F1 crossing the bottom face of the hole (see the top row of Fig. 18).  In the figure, each TCP has one or two branches, illustrating the flow from ill-bounded faces (second row) to options to resolve the ill-bounded faces (third row) and then to the resulting models from the respective options (bottom row).  For the first TCP, all the tested CAD software packages were able to resolve the ill-bounded faces successfully and gave the same modeling result as the proposed method.  However, for the second TCP, these packages were seen to yield different modeling results.  NX and SolidWorks failed to resolve the geometry-topology inconsistencies, producing an invalid model. SpaceClaim, Creo and Inventor were able to produce a valid model but the model shape change was not continuous. Only the proposed method successfully produced the desired result.

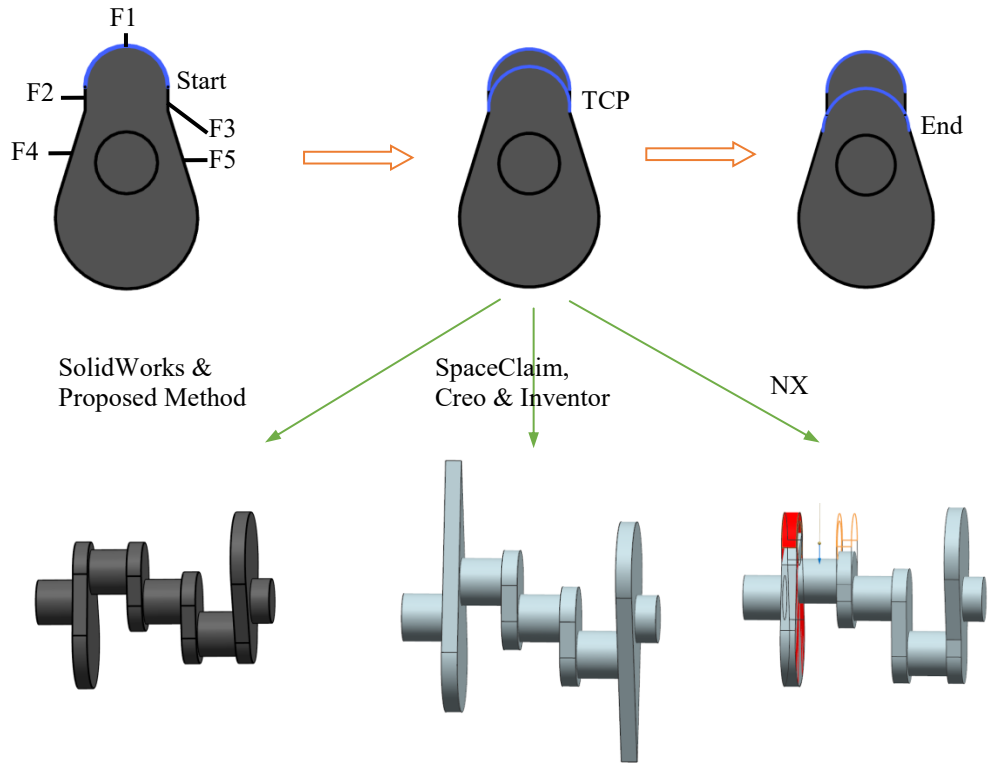

Figure 19   Comparison of modeling results for push-pulling the crank model.

By examining the difference between the modeling result of the proposed method and that of SpaceClaim, Creo and Inventor, it can be deducted that SpaceClaim, Creo and Inventor gave the undesired result likely due to a lack of the decomposition mechanism in their methods. The modeling result of SpaceClaim, Creo and Inventor possesses extra cuts (as circled in Fig. 18), which are very likely to be produced by the side faces F3 and F4 of the hole. However, there should not be extra intersections between faces F3, F4 and faces F5, F6 if model regeneration for the second TCP is performed with respect to the resulting model of the first TCP. In other words, it is impossible to get a resolution option from the ill-bounded faces in the left branch of the second TCP that can lead to the modeling result of SpaceClaim, Creo and Inventor. A plausible path that results in this modeling result is illustrated in the right branch of the second TCP. It should be noted that, even following the right branch, the desired modeling result can still be attained as indicated by the dashed arrow. However, it would be much harder to achieve the desired modeling result this way than employing the TCP based decomposition. This is because the more TCPs a push-pull move crosses, the less similar the final model topology is to the original one. It is thus much more difficult to come up with the desired model topology directly at the end location of the push-pull process.

The push-pull move in the third case study was purposely chosen to involve only one TCP so that Boolean operation based decision-making becomes dominant in the resolution of geometry-topology inconsistency. The TCP is when faces F2 and F3 are consumed with face F1 meeting faces F4 and F5 (the top row of Fig. 19). Illustrations of the ill-bounded faces and resolution options are omitted this time as they are relatively straightforward for this case. The five CAD software packages again produced inconsistent modeling results. NX failed to produce a valid model. SpaceClaim, Creo and Inventor output a valid model but the model shape change was not continuous. SolidWorks gave the same modeling result as the proposed method. The probable reason for the discontinuous modeling result of SpaceClaim, Creo and Inventor is that they simply removed face F1 with the consumed faces F2 and F3, resulting in an abrupt change of the model geometry. The proposed method was able to successfully produce a valid modeling result with a continuous model shape change by applying the principle of continuity which, for this case, simply removed the volume swept by face F1.

## 7. Conclusions

Push-pull direct modeling is available in some CAD software packages. However, it has been discovered that a robust method for push-pull direct modeling to consistently attain valid modeling results and continuous shape changes has not been reached yet. Further, the major source of the robustness issues is found to be the intrinsic ambiguity in resolving the geometry-topology inconsistency caused by the push-pull move. A new and systematic method has been presented in this paper to robustly resolve the geometry-topology inconsistency. The robustness is essentially achieved by applying the principle of continuity to guide the resolution of any inconsistent situation, which in turn was implemented via two modules: a TCP based decomposition module that tracks the TCPs along the push-pull process, and a Boolean operation based decision-making module that resolves inconsistency when crossing a TCP.

It should be noted that Boolean operations are quite compute-intensive and thus, the involved computational load may become significant when a good number of faces are push-pulled across a large and complex model. Parallel computing and localized Boolean operations maybe used to address this issue. It should also be noted that the proposed method assumes that the neighboring faces are stationary or move as a rigid with the push-pulled faces. There indeed exists another push-pull direct modeling mode: the neighboring faces move with the push-pulled face in a flexible manner so that the pre-edit connection continuity (e.g., $G^1$) between the face and the neighboring faces are to be maintained. For such cases, the presented method will fail to update the models. Push-pulling with the connection continuity is a challenging task because the neighboring faces' motions are governed by a system of nonlinear equations describing the continuity and there are no explicit expressions for these motions. Without knowing the motions explicitly, it is clearly difficult to predict/track the geometry-topology inconsistencies caused by the subjects of these motions. Addressing this challenge is among the direct modeling research studies to be carried out in our research group. Additionally, combining push-pull direct modeling with virtual reality techniques is of great interest for future studies.

## Acknowledgments

This work was in part funded by the Natural Sciences and Engineering Research Council of Canada (NSERC). We also thank the authors of the open source programming libraries: Open CASCADE, OpenFlipper and QT, as well as the authors of the parts used in our case studies.